\documentclass{owrart}

\usepackage{color}

\usepackage{amsthm}
 
\newtheorem{proposition}{Proposition}

\DeclareMathOperator{\res}{res}

\begin{document}


\begin{talk}[]{Daniela Cadamuro}
{Direct construction of pointlike observables in the Ising model}
{Cadamuro, Daniela}

\noindent

Relativistic quantum field theories are described by their set of local observables. These are linear bounded or unbounded operators associated with regions of Minkowski space. They form $\ast$-algebras that are expected to satisfy, e.g., the Haag-Kastler axioms, which are relevant to their interpretation as physical ``measurements''.

The problem of constructing models of quantum field theory, i.e., exhibiting algebras of local observables with such properties, is a notoriously hard task due to the complicated structure of local observables in the presence of interaction.

Quantum integrable models in $1+1$-dimensional Minkowski space are simplified models of interaction, rendering the mathematical structure of quantum field theory more accessible. In these models, the scattering of $n$ particles is the product of two particle scattering processes, namely the $S$-matrix is said to be ``factorizing'', a property connected to integrability. Examples include the Ising model, the $O(N)$ nonlinear sigma models and the Sine-Gordon model.

We are interested in studying the content of local observables in these theories. This can be investigated in various mathematical frameworks: as Wightman fields \cite{SW64}, as algebras of bounded operators \cite{Haa96}, or as closed operators affiliated with those algebras.
For example, the task of constructing the Wightman $n$-point functions in integrable models from a given $S$-matrix has been widely studied, see, e.g., \cite{Smi92}, but convergence of the associated series expansions has not been established so far, despite some progress \cite{BK04}.

An alternative approach considers fields localized in unbounded wedge-shaped regions as an intermediate step to the construction of sharply localized objects, which is handled indirectly \cite{Sch97,Lec08,Ala17,CT15}, thus avoiding explicit computation of pointlike fields. The existence proof of local observables is reduced to an abstract condition on the underlying wedge algebras. While the generators of the wedge algebras are explicitly known, the passage to the von Neumann algebras includes the weak limit points of this set. These limit points include the elements of local algebras, but of these much less is known.
Our task is to gain more information on the structure of these local observables, in the form of smeared pointlike quantum fields.  For this, we use a novel approach: instead of considering the $n$-point functions of pointlike fields and verifying the Wightman axioms, we establish them as closed operators affiliated with the local algebras above. 

In joint work with H.~Bostelmann  \cite{BC13} we characterize the local observables in terms of a family of coefficient functions $f_{m,n}^{[A]}$ in the following series expansion:
\begin{equation}\label{exp}
A = \sum_{m,n=0}^\infty \int \frac{d^m\pmb{\theta}d^n\pmb{\eta}}{m!n!} f_{m,n}^{[A]}(\pmb{\theta},\pmb{\eta}) z^\dagger(\theta_1)\cdots z^\dagger(\theta_m)z(\eta_1) \cdots z(\eta_n) ,
\end{equation}
where $z^\dagger,z$ are ``interacting'' creators and annihilators fulfilling a deformed version of the CCR relations which involves the scattering function. 

Due to the form of this expansion, local observables are defined as quadratic forms in a suitable class.
We denote  $\mathcal{H}^{\omega,f}$ the dense space of finite particle number states $\Psi$ fulfilling the condition $\|  e^{\omega(H/\mu)}\Psi \| < \infty$, where $H$ is the Hamiltonian, $\mu>0$ is the mass, and $\omega : [0, \infty) \to [0, \infty)$ is a function with the properties of \cite[Definition~2.1]{BC15}. In \eqref{exp}, $A$ is a quadratic form on $\mathcal{H}^{\omega,f} \times \mathcal{H}^{\omega,f}$ such that 
\begin{equation*}
\| Q_k A e^{- \omega(H/\mu)} Q_k \| + \| Q_k e^{-\omega(H/\mu)} A Q_k \| <   \infty
\end{equation*}
for any $k \in \mathbb{N}_0$, where $Q_k$ is the projector onto the space of $k$ or fewer particles. We denote this class of quadratic forms by $\mathcal{Q}^{\omega}$.

In order to characterize the coefficients $f_{m,n}^{[A]}$ in terms of the localization of $A$ in spacetime, we need a notion of locality which is adapted to quadratic forms in the class $\mathcal{Q}^{\omega}$:
We say that $A \in \mathcal{Q}^{\omega}$ is \emph{$\omega$-local} in the double cone $\mathcal{O}_{x,y} := \mathcal{W}_x \cap \mathcal{W}'_y$ (where $\mathcal{W}_x$ denotes the right wedge with edge at $x$ and $\mathcal{W}'_y$ the left wedge with edge at $y$, with $x$ to the left of $y$) if and only if $[A, \phi(f)] = [A, \phi'(g)] = 0$ for all $f \in \mathcal{D}^\omega(\mathcal{W}_y')$ and all $g \in \mathcal{D}^\omega(\mathcal{W}_x)$, as a relation in $\mathcal{Q}^{\omega}$. Here $\phi, \phi'$ are the left and right wedge-local fields, respectively, $\mathcal{D}^\omega(\mathcal{W}_x)$ is the space of smooth functions compactly supported in $\mathcal{W}_x$  with the property that $\theta \mapsto e^{\omega(\cosh\theta)} f^\pm(\theta)$ is bounded and square integrable ($f^\pm$ is positive and negative frequency part of the Fourier transform, respectively.)

The notion of $\omega$-locality is weaker than the usual notion of locality in the net of $C^\ast$-algebras $\mathcal{A}(\mathcal{O}_{x,y})$. It does not imply that $A$ commutes with unitary operators $e^{i\phi(f)^-}$, or with an element $B \in \mathcal{A}(\mathcal{W}_x)$: if $A$ is just a quadratic form, it would not be possible to write down these commutators in a meaningful way. We therefore clarify how $\omega$-locality is related to the usual locality \cite{BC18}:
\begin{proposition}\label{prop}
\quad 
\begin{enumerate}
\item[(i)] Let $A$ be a bounded operator; then $A$ is $\omega$-local in $\mathcal{O}_{x,y}$ for some $x,y \in \mathbb{R}^2$ if and only if $A \in \mathcal{A}(\mathcal{O}_{x,y})$.
\item[(ii)] Let $A$ be a closed operator with core $\mathcal{H}^{\omega,f}$, and $\mathcal{H}^{\omega,f} \subset \operatorname{dom} A ^\ast$. Suppose that
\begin{equation*}
\forall g \in \mathcal{D}^\omega_{\mathbb{R}}(\mathbb{R}^2) \; :\; \exp(i\phi(g)^-)\mathcal{H}^{\omega,f} \subset \operatorname{dom}A. \qquad  (\ast)
\end{equation*}
Then $A$ is $\omega$-local in $\mathcal{O}_{x,y}$ if and only if it is affiliated with $\mathcal{A}(\mathcal{O}_{x,y})$.
\item[(iii)] In the case $S = -1$, statement (ii) is true even without the condition $(\ast)$.
\end{enumerate}
\end{proposition}
This proposition gives criteria for affiliation of closed operators to local algebras, but in examples, closability of a quadratic form $A$ is difficult to characterize in terms of the coefficients in the expansion \eqref{exp}. Moreover, not much is known about the domain of the closed	 operator. We therefore look for sufficient (but not necessary) conditions that allow to apply Proposition~\ref{prop}. We will understand \eqref{exp} as an absolutely convergent sum on a certain domain, using summability conditions on the norms of the coefficients $f_{m,n}^{[A]}$. The following proposition provides a sufficient criterion for closability of $A$ as an operator:
\begin{proposition}\label{prop:closability}
Let $A \in \mathcal{Q}^{\omega}$. Suppose that for each fixed $n$,
\begin{equation*}
 \sum_{m=0}^\infty \frac{2^{m/2}}{\sqrt{m!}} \big( \| f^{[A]}_{m,n} \|^\omega_{m \times n} + \| f^{[A]}_{n,m}  \|^\omega_{n \times m}  \big) < \infty.
\end{equation*}
Then, $A$ extends to a closed operator $A^-$ with core $\mathcal{H}^{\omega,f}$, and $\mathcal{H}^{\omega,f} \subset \operatorname{dom}(A^-)^\ast$.
\end{proposition}
To apply Proposition~\ref{prop} we therefore need to fulfill the condition in Proposition~\ref{prop:closability} and to show $\omega$-locality of $A$. Hence, we formulate the $\omega$-locality condition in terms of properties of the functions $f_{m,n}^{[A]}$. This is the content of \cite[Theorem~5.4]{BC15}, which we summarize briefly: $A$ is localized in the standard double cone $\mathcal{O}_r$ of radius $r$ if and only if the coefficients $f_{m,n}^{[A]}$ are boundary values of meromorphic functions $(F_k)_{k=0}^\infty$ on $\mathbb{C}^k$ (with $k=m+n$) with a certain pole structure, which are $S$-symmetric, $S$-periodic, and fulfill certain bounds in the real and imaginary directions, depending on $\omega$ and $r$, and which fulfill the \emph{recursion relations}
%
\begin{equation*}
\res_{\zeta_n-\zeta_m = i \pi} F_{k}(\boldsymbol{\zeta})
= - \frac{1}{2\pi i }
\Big(\prod_{j=m}^{n} S(\zeta_j-\zeta_m) \Big)
\Big(1-\prod_{p=1}^{k} S(\zeta_m-\zeta_p) \Big)
F_{k-2}( \boldsymbol{\hat\zeta} ).
\end{equation*}
The problem is now to find examples of functions $(F_k)_{k=0}^\infty$ fulfilling the above conditions of $\omega$-locality and closability via Proposition~\ref{prop:closability}.
In the case $S=-1$ (Ising model) this is possible, and we aim at constructing a large enough set of observables so that they have the Reeh-Schlieder property. 

To that end, let $k\geq 0$, let $g \in \mathcal{D}(\mathcal{O}_r)$ with some $r>0$, and let $P$ be a symmetric Laurent polynomial of $2k$ variables. We define the  analytic functions
\begin{equation}
F_{2k}^{[2k,P,g]}(\pmb{\zeta}) := \tilde{g}(p(\pmb{\zeta})) P(e^{\pmb{\zeta}}) \sum_{\sigma \in \mathfrak{S}_{2k}} \operatorname{sign} \sigma \prod_{j=1}^k \sinh  \frac{\zeta_{\sigma(2j -1)} - \zeta_{\sigma(2j)}}{2},
\end{equation}
and $F^{[2k,P,g]}_j=0$ for $j \neq 2k$. For these the properties above hold with respect to this $r$ and for example with $\omega(p):= \ell \log(1+p)$ for some $\ell>0$.

Another example, involving the $F_j$ for odd $j$, is the non-terminating sequence
\begin{equation}
F_{2j+1}^{[1,P,g]}(\pmb{\zeta}) := \frac{1}{(2\pi i)^k} \tilde g(p(\pmb{\zeta})) P_{2j+1}(e^{\pmb{\zeta}}) \prod_{1 \leq \ell < r \leq 2j+1} \tanh \frac{ \zeta_{\ell} - \zeta_{r}}{2},
\end{equation}
where $g \in \mathcal{D}^\omega (\mathcal{O}_r )$, and $P=(P_{2j+1})_{j=0}^\infty$ are suitable chosen symmetric Laurent polynomials in $2j+1$ variables such that $P_{2j+1}(p,-p,\mathbf{q})= P_{2j -1}(\mathbf{q})$ (e.g., $P_{2j+1}(\pmb{x})=\sum_{\ell=1}^{2j+1}x_\ell^{2s+1}$, $s \in \mathbb{Z}$). We set $F_{2j}^{[1,P,g]}=0$. Also for these $F_j$, the properties above hold with respect to $r$ and $\omega(p)= p^\alpha$ with $\alpha \in (0, 1)$. 

Hence in both examples the associated quadratic form $A$ given by \eqref{exp} is $\omega$-local in the double cone $\mathcal{O}_r$. Additionally, the families of functions fulfill the condition of Proposition~\ref{prop:closability}, which implies that $A$ extends to a closed operator affiliated with the local algebras $\mathcal{A}(\mathcal{O}_r)$. This is in fact trivial for $(F^{[2k,P,g]}_j)_{j=0}^\infty$ as the sequence terminates; but for $(F^{[1,P,g]}_j)_{j=0}^\infty$, it involves careful norm estimates of a sequence of singular integral operators, as one is concerned precisely with the boundary values of the function at the poles of the hyperbolic tangent.

Further, we show that we find \emph{all} examples of pointlike local observables in the following sense. By choosing different polynomials $P$ we can generate a large set of observables which has the Reeh-Schlieder property, i.e., $\mathcal{Q}^{\omega}(\mathcal{O})\Omega$ is dense in the Hilbert space of the model, where $\mathcal{Q}^{\omega}(\mathcal{O})$ is our class of quadratic forms localized in $\mathcal{O}$. Even more, this $\mathcal{Q}^{\omega}(\mathcal{O})$ generates the local algebras in the sense that the net of algebras obtained from the spectral data of operators in $\mathcal{Q}^{\omega}(\mathcal{O})$ gives back the net $\mathcal{A}(\mathcal{O})$ after dualization.

\end{talk}

\end{document}